\documentclass[aps,prl,superscriptaddress,twocolumn,showpacs]{revtex4}
\usepackage{graphics}
\usepackage{epsfig}
\usepackage{color}
\usepackage{stackrel}
\usepackage{amsfonts}
\usepackage{wrapfig}
\usepackage{pstricks}
\usepackage{pst-node}
\usepackage{bm}
\usepackage{dcolumn}
\usepackage{multirow}
\usepackage{epstopdf}
\usepackage{subfigure}
\usepackage{amssymb}
\usepackage{amsmath}
\usepackage{commath}
\usepackage{graphicx,bm}
\usepackage{verbatim}
\usepackage{booktabs}
\usepackage[para]{threeparttable}
\usepackage{etoolbox}
\usepackage{lipsum}
\usepackage{marvosym}
\usepackage{wasysym}

\begin{document}
\title{Excitons in fractionally-filled moir\'{e} superlattices}

\author{Junghwan Kim}
\affiliation{Department of Electrical and Computer Engineering, University of Rochester, Rochester, New York 14627, USA}

\author{Hanan~Dery}
\altaffiliation{hanan.dery@rochester.edu}
\affiliation{Department of Electrical and Computer Engineering, University of Rochester, Rochester, New York 14627, USA}
\affiliation{Department of Physics and Astronomy, University of Rochester, Rochester, New York 14627, USA}

\author{Dinh~Van~Tuan}
\altaffiliation{vdinh@ur.rochester.edu}
\affiliation{Department of Electrical and Computer Engineering, University of Rochester, Rochester, New York 14627, USA}

\begin{abstract}
Long-range Coulomb forces give rise to correlated insulating states when charge particles populate a moir\'{e} superlattice at certain fractional filling factors. Such behavior is characterized by a broken translation symmetry wherein particles spontaneously form a Wigner crystal. Focusing on the experimental findings of Xu \textit{et al.} [Nature \textbf{587}, 214 (2020)], we present a theory that captures the correlated insulating state of a fractionally-filled moir\'{e} superlattice through the energy shift and change in oscillator strength of the exciton absorption resonance. The theory shows that the experimental findings can only be supported if the electrons reside in a charge-ordered state (i.e., electrons are not randomly distributed among the sites of the moir\'{e} superlattice). Furthermore, we explain why the energy shifts of exciton resonances are qualitatively different in cases that the superlattice is nearly empty compared with a superlattice whose sites are doubly occupied.
\end{abstract}
\maketitle

Correlated electrons states in moir\'{e} superlattices have recently become a focused research area with promising applications \cite{Nuckolls_NRM24,Andrei_NRM21,Du_Science23,Paik_AOP24,Ju_NRM24}. These  superlattices are created by stacking atomically thin graphene or transition-metal dichalcogenide (TMD) monolayers with a slight twist angle between them, resulting in a unique in-plane interference pattern \cite{Bistritzer_PNAS11,Wu_PRL17,Wu_PRL18,Wu_PRL19,Zhang_PRB20}. Since the lattice constant of the moir\'{e} superlattice is much larger than the lattice constants of atomically thin crystals, the edges of the conduction and valence bands of the underlying graphene or TMD monolayers split into flat energy minibands. The relatively large effective mass of electrostatically-doped electrons (or holes) in these flat minibands reduces the role of the kinetic energy and enhances the role of the Coulomb interaction. As a consequence, strongly correlated electron states emerge with exotic physical properties \cite{Tang_Nat20,Wang_NatMater20,Shimazaki_Nat20,Huang_NatPhys21,Jin_NatMater21,Ghiotto_Nat21,BenMhenni_arXiv24}, including Wigner crystals \cite{Zhou_Nature21,Wang_NatMat23}, Mott insulating states \cite{Li_Nat21,Regan_Nat20}, fractional quantum anomalous Hall effects \cite{Cai_Nature23,Zeng_Nature23,Park_Nature23,Xie_Nature21,Lu_Nature24,Xu_PRX23}, and superconducting states \cite{Cao_Nature18,Park_Nature21,Park_NatMater22}.

In many of these experiments, the correlated states are measured by optical spectroscopy of the exciton resonances through their energy shifts and changes in their oscillator strengths \cite{Regan_Nat20,Tang_Nat20,Shimazaki_Nat20,BenMhenni_arXiv24,Jin_NatMater21,Li_Nat21,Cai_Nature23,Zeng_Nature23,Park_Nature23,Wang_NatMat23,Zhou_Nature21}. The goal of the theory presented in this Letter is to show how the energy shifts and oscillator strengths depend on the fractional filling of the moir\'{e} superlattice, and to show that the experiments can only be explained if the electrons form a Wigner crystal. To streamline the discussion and directly compare the theory with experiment, we first summarize the key experimental results of Xu \textit{et al.} in Ref.~\cite{Xu_Nature20}, who have demonstrated that the exciton absorption spectrum becomes exceptionally rich when the moir\'{e} superlattice is filled with electrons or holes at various fractional fillings. After presenting these results, we will develop the theoretical model and will show that it fully supports and explains the experimental findings. 

\begin{figure*}
\includegraphics[width=18cm]{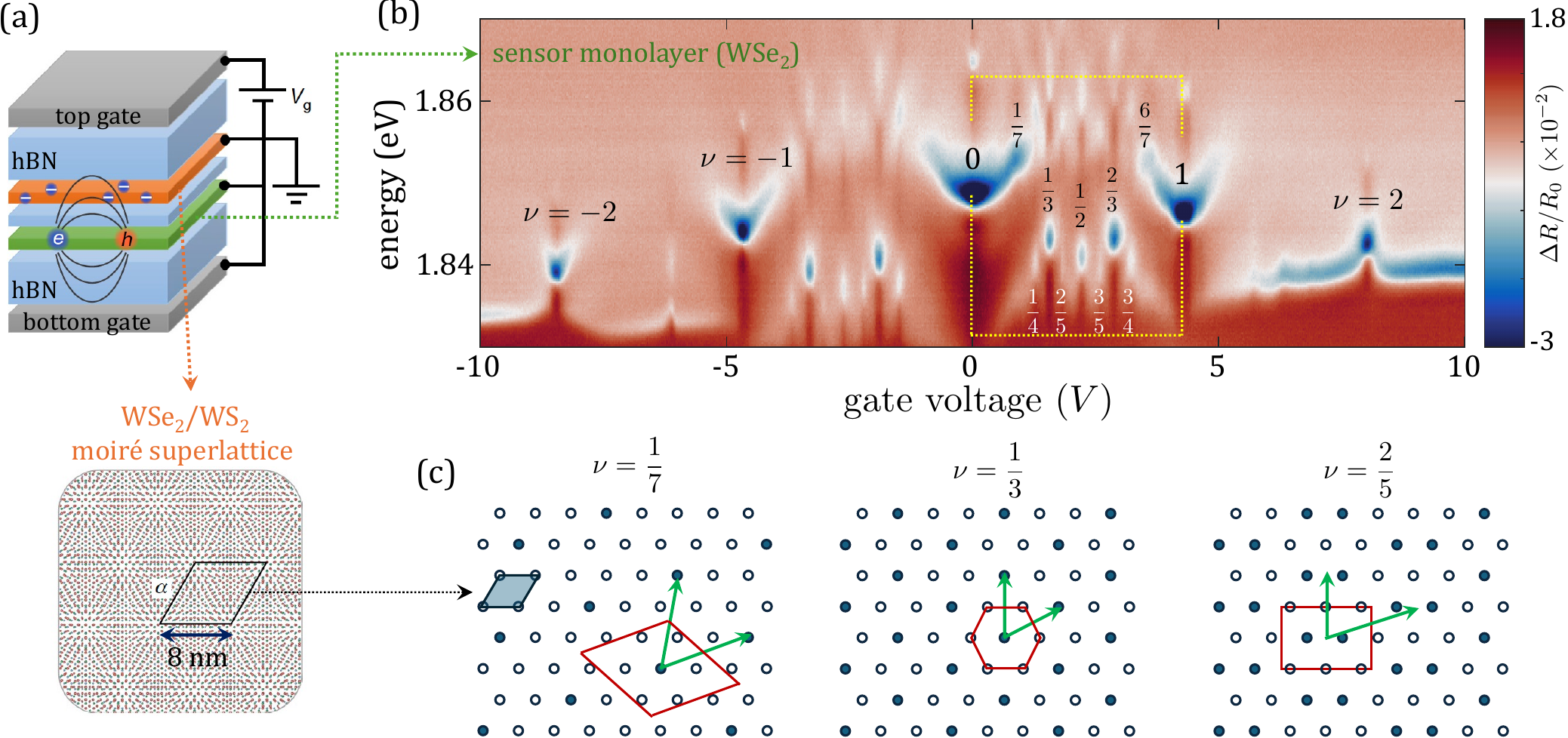}
 \caption{Experimental results of Ref.~\cite{Xu_Nature20}, provided by courtesy of Y. Xu, K. F. Mak and J. Shan. (a) Device structure and electric circuitry. The optically active regions of the stack are the sensor monolayer (charge neutral WSe$_2$),  separated by a thin hBN layer from an electrostatically-doped moir\'{e} superlattice (angle-aligned WSe$_2$/WS$_2$ bilayers; bottom scheme). (b) Gate-dependent reflection contrast at $T$$\,$$=$$\,$$1.6\,$K, shown in the spectral window of the 2$s$ exciton state in the sensor monolayer. The resonances are labeled by filling factors of the moir\'{e} superlattice at the corresponding gate voltages. The highlighted box between 0 and 4~V is analyzed in Fig.~\ref{fig:theory} where we compare these results with the theory. (c) Three examples of charge-order configurations at zero temperature of the fractionally-filled moir\'{e} superlattice. Filled and unfilled circles denote occupied and empty sites, respectively. Also shown are the unit cells and lattice vectors of each example.} \label{fig:exp} 
\end{figure*} 

Figure~\ref{fig:exp}(a) shows the device structure used in Ref.~\cite{Xu_Nature20}, where a WSe$_2$ sensor monolayer is placed a few nm below a moir\'{e} superlattice made of angle-aligned WSe$_2$/WS$_2$ bilayers (bottom panel), where a thin hexagonal boron nitride (hBN) layer is placed between the sensor and bilayers. The circuitry of the device guarantees that the sensor monolayer remains intrinsic and the electrostatic doping is only introduced in the bilayer system. The moir\'{e} potential lowers the optical gap in the WSe$_2$ part of the bilayer system by $\sim$50~meV compared with the optical gap of the isolated WSe$_2$ monolayer \cite{Xu_Nature20}. This difference allows one to unambiguously determine that the reflection contrast map in Fig.~\ref{fig:exp}(b) corresponds to the spectral region of the $2s$ exciton state in the sensor monolayer. The reason that the 2$s$ state has the most salient response to the moir\'{e} superlattice will become clear later. The goal of the model we are about to present is to reproduce the exciton resonances in Fig.~\ref{fig:exp}(b) for each of the fractionally-filled states both in terms of their energies and amplitudes.

We consider an exciton moving in a periodic moir\'{e} potential $V_M(\mathbf{R}, \mathbf{r})$ where $\mathbf{R}$ and $\mathbf{r}$ are the exciton's center-of-mass and relative motion position vectors, respectively. The exciton Hamiltonian reads
\begin{eqnarray}
H = - \frac{\hbar^2\nabla_\mathbf{R}^2}{2M} - \frac{\hbar^2\nabla_\mathbf{r}^2}{2\mu} + V(\mathbf{r}) + V_M(\mathbf{R}, \mathbf{r})\,, \label{eq:H}
\end{eqnarray}
where $V(\mathbf{r})$ is the electron-hole Coulomb interaction. $M=m_e + m_h$ and $\mu = m_e m_h /(m_e + m_h)$ are the translational and reduced masses of the exciton, where $m_{e(h)}$ is the effective mass of its electron (hole) component. The periodic potential satisfies $V_M(\mathbf{R}, \mathbf{r}) = V_M(\mathbf{R} + n_1\mathbf{R}_1 + n_2\mathbf{R}_2, \mathbf{r})$, where $n_{1(2)}$ are integers and $\mathbf{R}_{1(2)}$ are the basis lattice vectors, denoted by the arrows in Fig.~\ref{fig:exp}(c). The periodic potential can be expanded as a Fourier series,
\begin{eqnarray}
V_M(\mathbf{R}, \mathbf{r}) =  \sum_{\mathbf{G}} V_M(\mathbf{G}, \mathbf{r}) e^{i \mathbf{G}\cdot \mathbf{R}} \,,\, \label{eq:blochK}
\end{eqnarray} 
in which the sum runs over reciprocal lattice vectors $\mathbf{G}$. The Fourier components are
\begin{eqnarray}
V_M(\mathbf{G}, \mathbf{r}) = \frac{1}{A_{\nu}} \int_{A_\nu} V_M(\mathbf{R}, \mathbf{r}) e^{- i \mathbf{G}\cdot \mathbf{R}} d^2 \mathbf{R}\,,\, \label{eq:potFourier}
\end{eqnarray} 
where the integration region is the area of the unit cell ($A_\nu$). Examples of such unit cells at different fractional fillings are shown in Fig.~\ref{fig:exp}(c).

The periodicity of $V_M(\mathbf{R}, \mathbf{r})$ means that the exciton wavefunction has a Bloch form
\begin{eqnarray}
\Psi_{\mathbf{K}}(\mathbf{R}, \mathbf{r}) =  \sum_{\mathbf{G}} e^{i (\mathbf{K}+\mathbf{G})\cdot \mathbf{R}} u_\mathbf{K}(\mathbf{G}, \mathbf{r})\,,\, \label{eq:blochK}
\end{eqnarray} 
where the constant of motion $\mathbf{K}$ is the translational wavevector of the exciton. Projecting both sides of the Schr\"{o}dinger Equation $H \Psi_{\mathbf{K}}(\mathbf{R}, \mathbf{r}) = E_{\mathbf{K}}  \Psi_{\mathbf{K}}(\mathbf{R}, \mathbf{r})$ on $e^{i(\mathbf{K}+\mathbf{G})\cdot \mathbf{R}}$, we arrive at
\begin{eqnarray}
\!&\!&\!\! \! \! \! \! \! \left( \frac{\hbar^2(\mathbf{K}+\mathbf{G})^2}{2M} - \frac{\hbar^2\nabla_\mathbf{r}^2}{2\mu} + V(\mathbf{r})  \right) u_\mathbf{K}(\mathbf{G}, \mathbf{r}) \nonumber \\  \!&\!&\!\! \! \! \! \! \! \quad  + \sum_{\mathbf{G}'}  V_M(\mathbf{G} - \mathbf{G}', \mathbf{r})  u_\mathbf{K}(\mathbf{G}', \mathbf{r}) = E_{\mathbf{K}} u_\mathbf{K}(\mathbf{G}, \mathbf{r}). \,\,\,\,\,\,\,\,\,\,\label{eq:sec1}
\end{eqnarray} 
Hereafter, we focus on excitons in the light cone ($K=0$), which are the ones measured in Fig.~\ref{fig:exp}(b), and omit the wavevector subscript for brevity. To solve Eq.~(\ref{eq:sec1}), we expand the Bloch function as
\begin{eqnarray}
u(\mathbf{G}, \mathbf{r}) = \sum_{\alpha} C_{\alpha}(\mathbf{G}) \phi_\alpha(\mathbf{r}) \equiv  \sum_{\alpha} C_{\alpha}(\mathbf{G}) | \alpha \rangle  \,,\, \label{eq:bloch}
\end{eqnarray} 
where the sum runs over a complete set of states $\phi_\alpha(\mathbf{r})$ with $\alpha = \{ 1s,\,2s,\,2p^{\pm},...\}$ from the solutions of
\begin{eqnarray}
\left[ V(\mathbf{r}) - \frac{\hbar^2\nabla_\mathbf{r}^2}{2\mu} \right] \phi_\alpha(\mathbf{r}) = \varepsilon_\alpha \phi_\alpha(\mathbf{r})  \,.\, \label{eq:Hr}
\end{eqnarray}
Substituting Eqs.~(\ref{eq:bloch})-(\ref{eq:Hr}) in (\ref{eq:sec1}), we arrive at a secular equation that can be solved through matrix inversion 
\begin{eqnarray}
\!\!&\!\!&\!\!\!\!\!\! \sum_{\mathbf{G}',\beta} \left[ \!\left( \!\frac{\hbar^2\mathbf{G}^2}{2M} + \varepsilon_\alpha \!\right) \! \delta_{\mathbf{G},\mathbf{G}'} \delta_{\alpha,\beta} + V_{\alpha,\beta}(\mathbf{G} - \mathbf{G}') \right] \!C_{\beta}(\mathbf{G}')  \nonumber \\ 
&\,& \quad  = E C_{\alpha}(\mathbf{G})\,\,, \,\,\,\,\,\,\,\,\,\,\, \label{eq:sec}
\end{eqnarray}
where 
\begin{eqnarray}
V_{\alpha,\beta}(\mathbf{G}-\mathbf{G}') = \langle \beta | V_M(\mathbf{G}-\mathbf{G}',\mathbf{r})| \alpha \rangle
\,\, .\,\,\,\,\,\,\,\, \label{eq:Vab}
\end{eqnarray}
To evaluate these matrix elements we consider a periodic potential of the form
\begin{eqnarray}
V_M(\mathbf{R}, \mathbf{r}) &=&  \frac{e^2}{\epsilon} \sum_{\mathbf{R}_M}   \Bigg( \frac{1}{\sqrt{ d^2 + (\mathbf{R} + \frac{\mathbf{r}}{2} - \mathbf{R}_M)^2 }}  \nonumber \\ 
&\,& \quad \quad \quad -  \frac{1}{\sqrt{ d^2 + (\mathbf{R} - \frac{\mathbf{r}}{2} - \mathbf{R}_M)^2 }}\Bigg)\,\, .\,\,\,\,\,\,\,\,\,\,\, \label{eq:VRr}
\end{eqnarray}
$\epsilon$ is the effective dielectric constant of the environment, $\mathbf{R} + \mathbf{r}/2$ is the position of the electron component in the exciton, and $\mathbf{R} - \mathbf{r}/2$ is that of the hole component. $d$ is the distance between the plane at which the exciton resides in the sensor monolayer and the plane at which electrons reside in the moir\'{e} superlattice at positions $\mathbf{R}_M$. Substituting Eq.~(\ref{eq:VRr}) in Eqs.~(\ref{eq:potFourier}) and (\ref{eq:Vab}), we get after some algebra that the matrix elements of the periodic potential read 
\begin{eqnarray}
V_{\alpha,\beta}(\widetilde{\mathbf{G}}) &=&  \frac{2\pi e^2}{A_\nu \epsilon} \frac{e^{-d\widetilde{G}}}{\widetilde{G}}  \Bigg[ 2i \sum_{\mathbf{R}'_M }e^{-i\widetilde{\mathbf{G}} \cdot \mathbf{R}'_M}\Bigg] \mathcal{M}_{\alpha,\beta}(\widetilde{\mathbf{G}})\,,\,\,\,\,\,\,\,\,\,\,\,\,\,\,
\label{eq:Vab1}
\end{eqnarray}
where $\widetilde{\mathbf{G}} \equiv \mathbf{G}-\mathbf{G}'$. The sum runs over electrons positions in one unit cell $\mathbf{R}'_M$ (i.e., the filled circles in a unit cell in Fig.~\ref{fig:exp}(c)), and 
\begin{eqnarray} 
\mathcal{M}_{\alpha,\beta}(\widetilde{\mathbf{G}}) =  \int  \phi_\alpha^\ast(\mathbf{r}) \sin\Big(\tfrac{1}{2} \widetilde{\mathbf{G}}\cdot\mathbf{r} \Big) \phi_\beta(\mathbf{r}) d^2 \mathbf{r}\,.
\label{eq:Mab}
\end{eqnarray}
The odd sine function implies that $\mathcal{M}_{\alpha,\beta}(\widetilde{\mathbf{G}}) \neq 0$ only if $\alpha$ and $\beta$ have different parity. Consequently, the oscillator strengths and energies of the $s$-states, whose properties are probed in optical reflectance experiments, change because they are mixed through the moir\'{e} potential with exciton states of $\{p,\,d,\,...\}$ characters \cite{Lu_PRB24}. Given that $|\varepsilon_{2s} - \varepsilon_{2p^\pm}| \sim 10$~meV whereas $| \varepsilon_{1s} - \varepsilon_{2p^\pm}|\gtrsim 120$~meV in hBN-encapsulated TMD monolayers \cite{Zhu_PRL23,Donck_PRB19,Wang_PRL15,Yong_NatMater19}, the 2$s$ state exhibits a stronger response to the presence of electrons in the moir\'{e} superlattice in these type of experiments \cite{Xu_Nature20,Hu_Science23,He_NatMater24}. Whereas electrons in the moir\'{e} superlattice can also strongly mix $s$-state excitons at higher energies (3$s$, 4$s$,...) with exciton states at nearby energies (3$p^\pm$, 3$d^\pm$, 4$p^\pm$...), the detection of these larger-radii excitons is difficult on accounts of their weaker oscillator strength \cite{Stier_PRL18}.

The calculation results we are about to present only consider the small subspace of $2s$ and $2p^{\pm}$ hydrogen-like exciton states,
\begin{eqnarray} 
\phi_{2s}(\mathbf{r}) &=& \frac{1} {\sqrt{6\pi}} \, \frac{r_{0}- r }{r_{0}^2} \, e^{- r / 2r_{0}}\,,\nonumber \\
\phi_{2p^\pm}(\mathbf{r}) &=& \frac{e^{\pm i \theta}} {\sqrt{12\pi}} \, \frac{r}{r_{0}^2} \, e^{- r / 2r_{0}},
\label{eq:sp}
\end{eqnarray}
where $r_{0}=3\epsilon\hbar^2 / 4\mu e^2 $ and $\mathbf{r} = (r\cos \theta,\, r\sin \theta)$ \cite{Yang_PRA91}. As we will show, these three orbitals are sufficient to reproduce the experimental results of Fig.~\ref{fig:exp}. Substituting Eq.~(\ref{eq:sp}) in (\ref{eq:Mab}), we get $\mathcal{M}_{2s,2s}=\mathcal{M}_{2p^\pm,2p^\pm}=\mathcal{M}_{2p^\pm,2p^\mp}=0$ and 
\begin{eqnarray} 
\mathcal{M}_{2s,2p^\pm}(\widetilde{\mathbf{G}}) &=& 16 \sqrt{2} \, e^{\pm i\varphi}  \frac{ (r_0^2 \widetilde{G}^2 - 6) r_0 \widetilde{G} } {(r_0^2 \widetilde{G}^2 + 4)^{7/2}} \,,\,
\label{eq:Msp}
\end{eqnarray}
where $\widetilde{\mathbf{G}} = (\widetilde{G}\cos \varphi,\,\widetilde{G}\sin \varphi)$.

To compare the model with experimental findings, we solve Eq.~(\ref{eq:sec}) with the help of Eqs.~(\ref{eq:Vab1}) and (\ref{eq:Msp}) for various fractional filling factors $\nu$. The parameters that depend on $\nu$ are the unit cell area ($A_\nu$), electron positions in the unit cell ($\mathbf{R}'_M$), and the reciprocal lattice vectors $\mathbf{G}=\ell_1\mathbf{G}_{1} + \ell_2 \mathbf{G}_{2}$ through the basis vectors $\mathbf{G}_{1(2)}$. The number of reciprocal lattice vectors in the simulations is defined by all integers in the range $-N_G \leq \ell_{1},\ell_2 \leq N_G$. To simulate a disordered distribution of electrons, we consider a supercell that includes 144 moir\'{e} unit cells, and randomly choose $\nu$ of these unit cells to be filled (e.g., filling 48 of the 144 cells when $\nu=1/3$). The Supplemental Material includes technical details on the simulations in the ordered and disordered configurations for each case of $\nu$ that is covered in this work \cite{supp}.  

\begin{figure*}
\includegraphics[width=18cm]{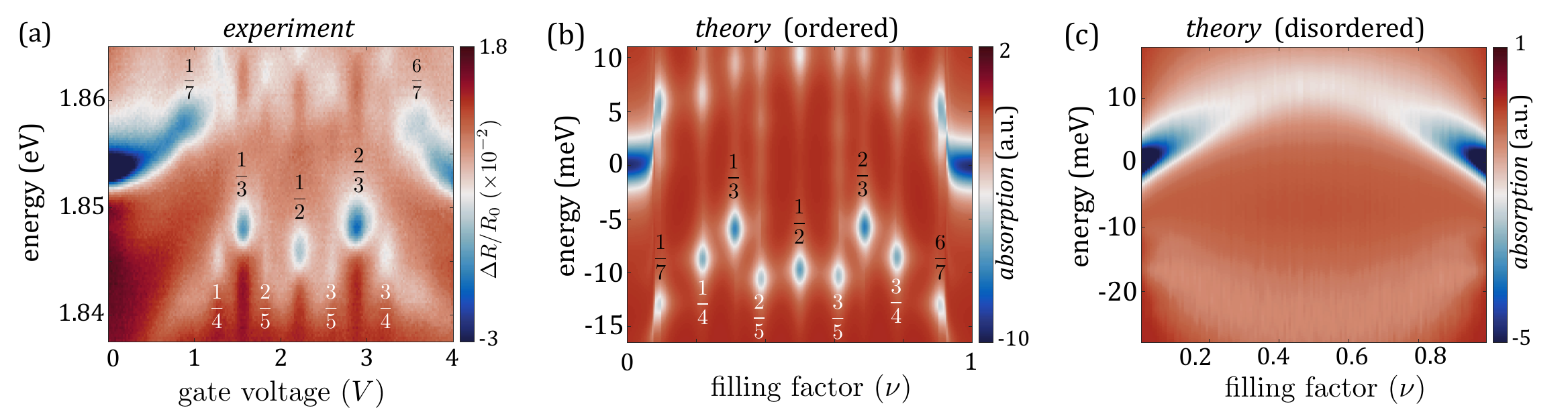}
 \caption{Comparing the measurement results with theory. (a) Experiment: Magnified view of the reflectance spectra in the highlighted box of Fig.~\ref{fig:exp}(b). (b) Theory: Optical absorption in the sensor WSe$_2$ monolayer, calculated by assuming an ordered-state of the electrons in the moir\'{e} superlattice (e.g., Fig.~\ref{fig:exp}(c)). The reference  level (zero energy) corresponds to the resonance energy of the $2s$ exciton state when the adjacent moir\'{e} superlattice is empty. (c) The same as in (b), but by assuming a disordered state of the electrons in the moir\'{e} superlattice.} \label{fig:theory}
\end{figure*}

The parameters used in the simulations are as follows. The translation and reduced masses of the exciton are $M=0.65m_0$ and $\mu=0.16m_0$, respectively, calculated by using $m_e=0.29m_0$ and $m_h=0.36m_0$ \cite{Kormanyos_2DMater15}. The energy of the 2$s$ state lies 10~meV above that of the degenerate ${2p^\pm}$ states  \cite{Zhu_PRL23,Donck_PRB19}. The lattice constant of the moir\'{e} superlattice is 8~nm \cite{Xu_Nature20}, the distance between the planes of the exciton and superlattice is $d=4$~nm, and $N_G = 8$ ($N_G = 13$) is found sufficient to achieve converged results in the ordered (disordered) configurations \cite{supp}. Finally, $r_{0}=3\epsilon\hbar^2 / 4\mu e^2 = 1.24$~nm is calculated by using $\epsilon=5$, chosen such that the relation between $r_0$ and the root mean square radius of the $2s$ exciton state, $\sqrt{\langle r^2 \rangle_{2s}} = \sqrt{\int d^2 \mathbf{r} \,r^2 | \phi_{2s}(\mathbf{r}) |^2} = \sqrt{26}r_0$, agrees with the experimental result $\sqrt{\langle r^2 \rangle_{2s}} = 6.6$~nm \cite{Stier_PRL18}.

After solving Eq.~(\ref{eq:sec}), we substitute Eq.~(\ref{eq:bloch}) in Eq.~(\ref{eq:blochK}) for the case of an exciton in the light cone ($K=0$), and evaluate the absorption profile from 
\begin{eqnarray} 
\alpha_\nu(\hbar \omega) &\propto& \sum_j \frac{\left|\int_{A_{\nu}} \Psi_{\nu,j}(\mathbf{R}, r) d^2\mathbf{R}\right|_{r=0}^2}{(\hbar \omega - E_{\nu,j})^2 + \delta^2} \nonumber \\
&\propto& \sum_j  \frac{ |C_{\nu,j,2s}(G=0)|^2}{(\hbar \omega - E_{\nu,j})^2 + \delta^2} \,.
\label{eq:abs}
\end{eqnarray}
The energy broadening parameter  is $\delta = 1$~meV, and $j$ runs over the eigenstates of Eq.~(\ref{eq:sec}). The wavefunction in the first line is evaluated at $r=0$ (the hole and electron positions have to overlap for linear absorption to take place). After integrating the translational position vector over the area  of the unit cell, we are left only with $G=0$ components of the $2s$ state ($\phi_{2p^{\pm}}(r=0)=0$).

Figure~\ref{fig:theory} shows the experimental results with the calculated absorption profile from Eq.~(\ref{eq:abs}). Figure~\ref{fig:theory}(a)  is a magnified view of the reflectance spectra in the highlighted box of Fig.~\ref{fig:exp}(b). Figures~\ref{fig:theory}(b) and (c) show the theory results when electrons in the moir\'{e} superlattice are charge-ordered and disordered, respectively. The results in Figs.~\ref{fig:theory}(a) and (b) match both in terms of the resonance energies and oscillator strengths. With respect to the strongest 2$s$ resonances at $\nu=0$ and 1, both experiment and theory show that the next strongest  resonances are at $\nu=1/3$ and 2/3 and they emerge $\sim$5~meV below the one at $\nu=0$. The next strongest resonance appears at $\nu=1/2$ and it emerges $\sim$9~meV below the one at $\nu=0$. Next in amplitudes are the resonances at $\nu=1/4$ and 3/4 that emerge $\sim$8~meV below in theory versus $\sim$10~meV in experiment, followed by the ones at $\nu=2/5$ and 3/5 that emerge $\sim$10~meV below in both theory and experiment. The weakest resonance emerge 13 meV below at $\nu=1/7$ and 6/7, in which case there is also a stronger resonance next to the one at $\nu=0$ in both experiment and theory. All in all, these results support the conclusion that particles of the moir\'{e} superlattice in Ref.~\cite{Xu_Nature20} form a Wigner crystal. 

This conclusion is further reinforced by the simulated absorption profile when assuming disordered configuration. The calculated absorption map in Fig.~\ref{fig:theory}(c) is completely different than the one seen in experiment, and this conclusion remains valid whether we average the absorption maps of many random distributions of electrons in the moir\'{e} superlattice or consider the absorption profile of a certain random distribution (see Supplemental Material). The absorption map in Fig.~\ref{fig:theory}(c) shows two energy `bands', where the absorption is stronger at the higher energy around zero energy, which blueshifts  continuously from $\nu=0$  to $\nu=1/2$ (or from $\nu=1$ to $\nu=1/2$). The second band with weaker absorption emerges $\sim$20~meV below. This band is reminiscent of an impurity band in doped semiconductors \cite{Stern_PR55,Conwell_PR56,Matsubara_PTP61}, where here it is an exciton that becomes localized next to disorder centers.

Finally, we focus on the energy shifts of the resonances next to integer fillings, seen both in the calculated absorption maps and in the measured reflection contrast map. Focusing on the larger reflection contrast map in Fig.~\ref{fig:exp}(b), we notice that the resonances near $\nu=0$ and $\nu=1$ blueshift in energy as we depart from integer filling, whereas the resonance near $\nu=2$ redshifts. These opposite trends depend on whether the electrons in the moir\'{e} superlattice are localized or itinerant. When the electrons are localized and their density increases, then in addition to the extended exciton states near or above zero energy, excitons states develop at lower energies (i.e., bound to moir\'{e} cells filled with electrons). For an exciton to remain extended, its energy has to increase (blueshift) in order to be orthogonal to the lower exciton states \cite{VanTuan_PRB23L,VanTuan_PRB23}. The oscillator strength is progressively transferred from the extended excitons to the localized ones, and this effect becomes strong enough to become observable when $\nu=1/7$, as shown in Fig.~\ref{fig:theory}. The case of $\nu = 6/7$ is equivalent by considering the empty moir\'{e} cells as holes. The case near $\nu=2$ is different because electrons are no longer localized, and they act to screen the Coulomb interaction. The resulting effect is energy redshift of the resonance because of the interplay between screening-induced optical gap renormalization and reduced binding energy of the exciton \cite{Mhenni_ACSNano25,VanTuan_PRB24,Dery_arXiv24}.

In closing, we have presented a theory that studies how the resonance energy and oscillator strength of excitons depend on the fractionally-filled state of a moir\'{e} superlattice.  Evident differences are found between cases in which the electrons form a Wigner crystal compared with cases in which the electrons are randomly distributed among the unit cells of the moir\'{e} superlattice. Comparing the calculated results with recent experiments, we conclude that charge-order states are responsible to the observed behavior in these experiments. Future experiments can further investigate the mixing between the $s$ and $p$ states of the exciton through the moir\'{e} potential \cite{Lu_PRB24}, making use of out-of-plane magnetic fields or in-plane electric fields to split the $p^{\pm}$ states, acting as knobs to control the polarization, amplitude and energy of the exciton resonance.  
%

\acknowledgments{We are indebted to Jie Shan, Kin Fai Mak, and Yang Xu for sharing the experimental data and for fruitful discussions. This work is supported by the Department of Energy, Basic Energy Sciences, Division of Materials Sciences and Engineering under Award No. DE-SC0014349.} 


\clearpage

\begin{widetext}
\section{Lattice information}

In this section, we specify the unit cells and lattice vectors of several fractional filled moir\'e superlattices. In a two-dimensional Bravais lattice, the crystal structure is defined by two real-space lattice vectors, which we denote as $\{\mathbf{R}_1,\,\mathbf{R}_2\}$.
These vectors span the 2D lattice in real space. The reciprocal lattice vectors, \(\mathbf{G}_1\) and \(\mathbf{G}_2\), are defined by the condition
\[
\mathbf{G}_i \cdot \mathbf{R}_j \;=\; 2\pi \,\delta_{i,j}
\quad
(i,j=1,2),
\]
where \(\delta_{i,j}\) is the Kronecker delta.

\begin{figure}[ht]
 \centering
    \includegraphics[width=0.9\textwidth]{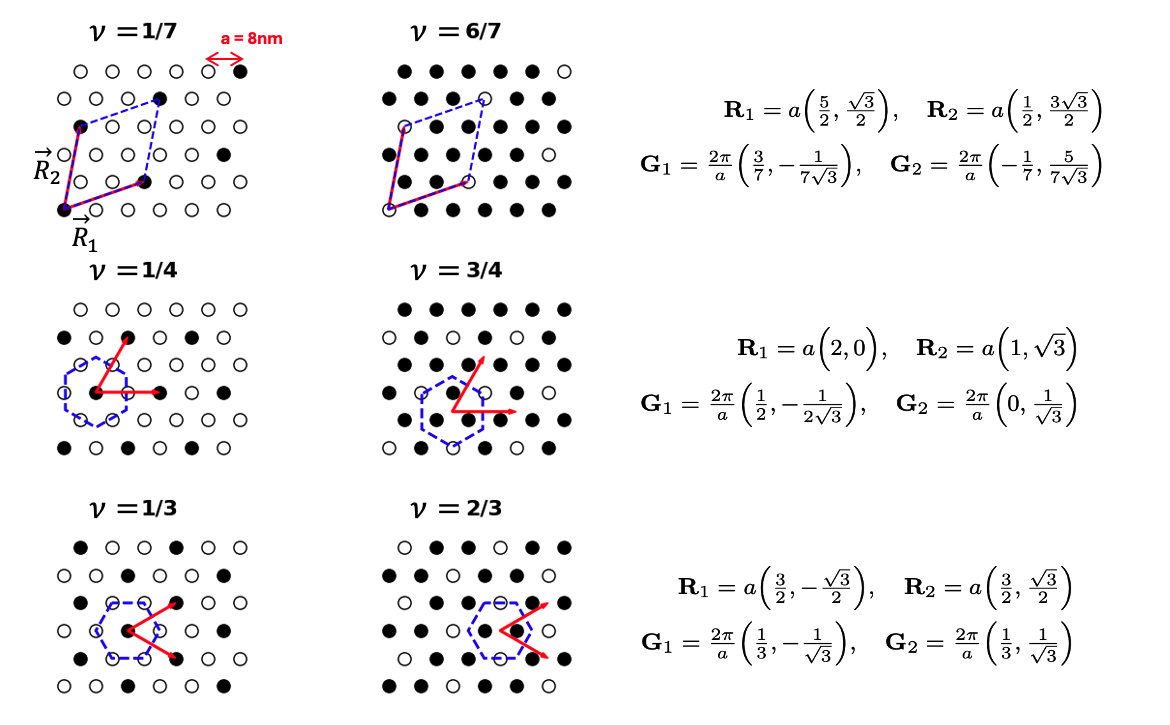} 
\caption{Lattice configurations of fractionally-filled moir\'{e} superlattices corresponding to various filling factors. Filled (empty) circles denote occupied (empty) sites. Red arrows indicate the real-space lattice vectors, $\{\mathbf{R}_1, \mathbf{R}_2\}$, and dashed blue lines mark the unit cell of each filling factor. The corresponding real-space and reciprocal lattice vectors are listed on the right of each panel. Complementary filling factors, $\nu$ and 1-$\nu$, share the same real-space and reciprocal lattice vectors. The moir\'{e} lattice constant is $a = 8$~nm, as indicated in the panel of $\nu=1/7$.}
\label{fig:S1}
\end{figure}

\begin{figure}[ht]
 \centering
    \hspace{-15mm}
    \includegraphics[width=0.9\textwidth]{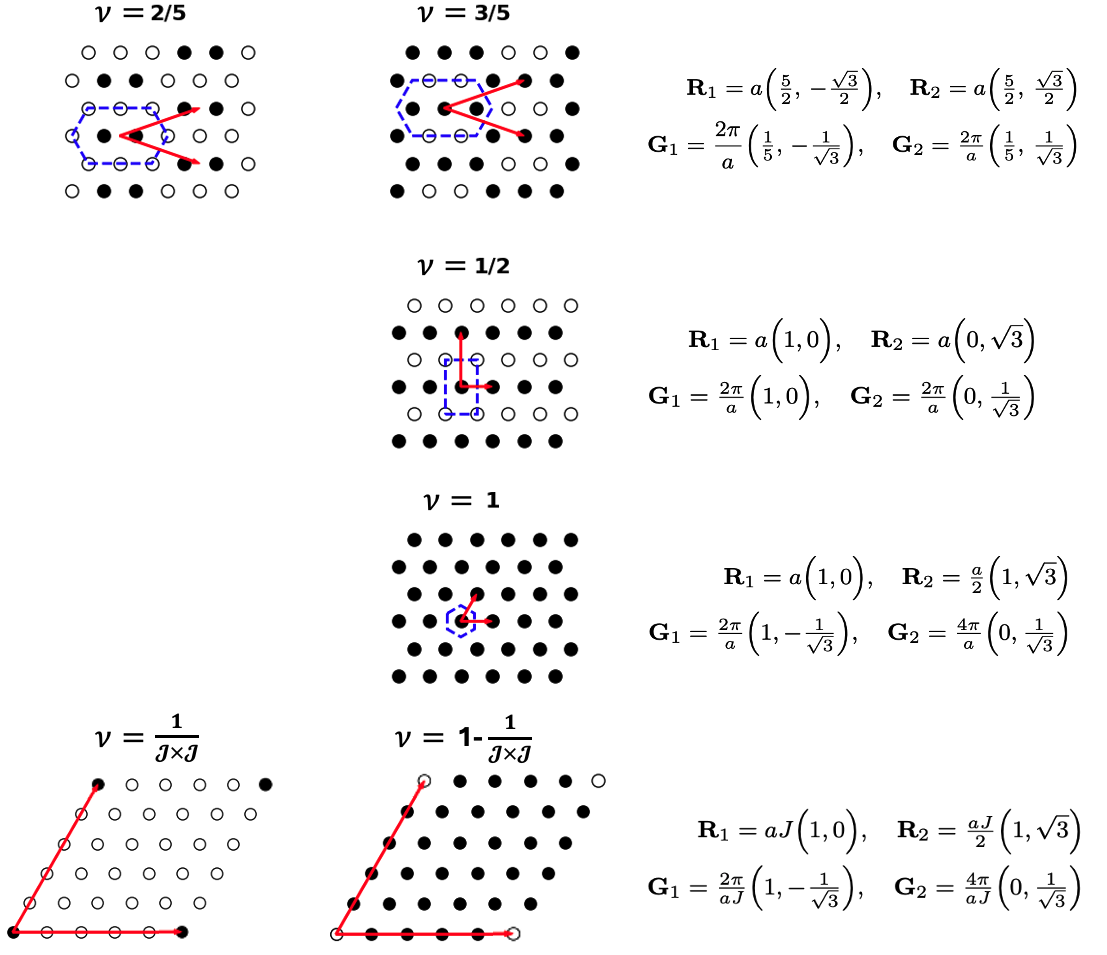}
    \caption{%
Lattice configurations of fractionally filled moir\'{e} superlattices at various filling factors.
In the bottom row, a generalized $J \times J$ supercell is shown, illustrating how nearly empty ($\nu \sim 0$) or nearly full ($\nu \sim 1$) fillings can be constructed.
}
\label{fig:S2}
\end{figure}

As shown in Fig.~S1, complementary fillings ($\nu$ and $1-\nu$) have the same unit cells, real-space and reciprocal lattice vectors. The only difference between them is the number of electrons per unit cell. The difference in electron count between $\nu$ and $1-\nu$ leads to different Coulomb interaction between electrons in the moir\'{e} lattice and the exciton, calculated by the form factor, $F_{\nu}(\widetilde{\mathbf{G}}) \;\equiv\; \sum_{\mathbf{R}_\mathrm{M}} e^{ \,i\,\widetilde{\mathbf{G}}\cdot\mathbf{R}_\mathrm{M}}$, which is discussed in the main paper and elaborated on below. The unit cell area of each filling factor is $A_{\nu}$ = $\nu_d{A_1}$, where $A_1 = \tfrac{\sqrt{3}}{2}a^2$ is the unit cell area for $\nu = 1$ and $\nu_d$ is an integer  given by the denominator of the fraction $\nu$.

\vspace{2mm}

By extending the analogy between $\nu$ and $1-\nu$, one can also describe filling factors close to 0 or 1 by enlarging the unit cell and then placing only one electron in it (for $\nu \approx 0$), or remove one electron from a fully filled cell (for $\nu \approx 1$). For example, if we consider a $J \times J$ supercell for $J=10$ where each site is singly occupied at $\nu=1$, then removing a single electron would give a filling factor
\[
\frac{J^2 - 1}{\,J^2\,} \;=\; 1 \;-\;\frac{1}{\,J^2\,} = 0.99
\]
whereas keeping only a single electron in an otherwise empty $J \times J$ supercell results in a filling factor close to zero $1/J^2 = 0.01$. In both cases, the lattice vectors and supercell geometry remain the same; only the electron occupancy changes. This perspective makes it easier to simulate nearly empty (or nearly full) moir\'{e} lattices using a common framework.

\subsection{Derivation of form factors, {$F_{\nu}(\widetilde{\mathbf{G}})$} and examples}

In the following, we specify the form factors, $F_{\nu}(\widetilde{\mathbf{G}})$ for various fractional filling factors $\nu$. Our goal is to show how to simplify the external Coulomb potential matrix elements by appropriately choosing the origin of coordinates. We then provide step-by-step calculations of the phase sums
\begin{equation}
   F_{\nu}(\widetilde{\mathbf{G}}) \;\equiv\; \sum_{\mathbf{R}_\mathrm{M}\in A_{\nu}} e^{-\,i\,\widetilde{\mathbf{G}}\cdot \mathbf{R}_\mathrm{M}} \;=\; 
   \sum_{m=1}^{N_\nu}\! e^{-\,i\,\widetilde{\mathbf{G}}\cdot \mathbf{R}^m_{\mathrm{M}}},
   \label{eq:Fnu}
\end{equation}
where $N_\nu$ is the number of electrons in each unit cell for filling factor $\nu$. The sum runs over the occupied electron sites within one \textit{unit cell} of the moir\'{e} superlattice. In the main paper, expressions of the form factor arise when we derive matrix elements of the periodic potential. This treatment illustrates how fractionally filled moir\'{e} superlattices can be analyzed, enabling us to track how exciton energies and oscillator strengths depend on electron arrangements in real space. A recurring important idea in this analysis is that an appropriate choice of the origin for each fractional filling factor $\nu$ can simplify the real or imaginary phase factors in the form factors $F_{\nu}(\widetilde{\mathbf{G}})$. We show such examples for $\nu = 1/2,\;2/3,\;1/3,\;3/4,\;1/4,\;1,\;6/7$, and so on. The results confirm that in many cases one can eliminate net phase factors entirely, while in other cases only a pure overall phase remains. Nevertheless, these phase conventions do not affect the Hermiticity of the resulting Coulomb matrix.

\subsubsection*{\textbf{Filling factors with one electron in the unit cell} $(\nu=\{ 1/J^2,\,1/7,\,1/4,\,1/3,\,1/2,\,1\})$}
The unit cell has only one electron, and therefore, we choose the origin as the position of this electron, meaning that $\mathbf{R}^1_{M} = (0,0)$ and
\begin{equation}
F_{\nu}(\widetilde{\mathbf{G}}) \;=\; 1.
\end{equation}

\subsubsection*{\textbf{Filling factor $\nu = 2/3$}}

The unit cell has two electrons, and we choose the origin at the midpoint of the two electrons, meaning that $\{\mathbf{R}_M^{1,2}\} = \bigl\{\bigl(-\tfrac{a}{2},\,0\bigr),\,\bigl(\tfrac{a}{2},\,0\bigr)\bigr\}$ and 
\begin{align}
F_{\nu=2/3}(\widetilde{\mathbf{G}})
&=\; \Bigr[\exp\!\Bigl(i\,\tfrac{\pi}{3}\,(i_1 + i_2)\Bigr)
\;+\;
\exp\!\Bigl(-\,i\,\tfrac{\pi}{3}\,(i_1 + i_2)\Bigr)\Bigr] \nonumber\\
&=\; 2\cos\!\Bigl(\tfrac{\pi}{3}\,(i_1 + i_2)\Bigr)
,
\end{align}
where
\begin{equation}
\widetilde{\mathbf{G}} \;=\; i_1\,\mathbf{{G}}_1 + i_2\,\mathbf{{G}}_2 
\;=\;
\frac{2\pi}{a}\Bigl(\frac{i_1 + i_2}{3},\,\frac{i_2 - i_1}{\sqrt{3}}\Bigr), \label{eq:Gt}
\end{equation}
and $\{i_1,i_2\}$ are integers.

\subsubsection*{\textbf{Filling factor  {$\nu = 3/4$}}}

The unit cell has three electrons forming an equilateral triangle, and we choose the origin at the center of that triangle, meaning that
\[
\{\mathbf{R}_M^{1,2,3}\}
=\Bigl\{\bigl(-\tfrac{a}{2},\,-\tfrac{a}{2\sqrt{3}}\bigr),
\bigl(\tfrac{a}{2},\,-\tfrac{a}{2\sqrt{3}}\bigr),
\bigl(0,\,\tfrac{a}{\sqrt{3}}\bigr)\Bigr\},
\]
and
\begin{align}
F_{\nu=3/4}(\widetilde{\mathbf{G}}) &=
\Bigl[\exp\!\Bigl(i\,\tfrac{\pi}{3}(i_1 + i_2)\Bigr)
+ \exp\!\Bigl(-\,i\,\tfrac{\pi}{3}(2i_1 - i_2)\,\Bigr)
+ \exp\!\Bigl(-\,i\,\tfrac{\pi}{3}(2i_2 - i_1)\Bigr)\Bigr],
\end{align}
where 
\begin{equation}
\mathbf{\widetilde{G}} \;=\; i_1\,\mathbf{{G}}_1 + i_2\,\mathbf{{G}}_2 
\;=\;
\frac{2\pi}{a}\Bigl(\tfrac{i_1}{2} \;,\; \tfrac{2\,i_2 - i_1}{2\,\sqrt{3}}\Bigr)\\[15pt].
\end{equation}


\subsubsection*{\textbf{Filling factors $\nu = 2/5$ and {$\nu = 3/5$}}}
At $\nu=3/5$, there are three electrons in the unit cell at
\[
\{\mathbf{R}_M^{1,2,3}\}
=\Bigl\{\bigl(0,\,0\bigr),
\bigl(a,\,0\bigr),
\bigl(4a,\,0\bigr)\Bigr\}\,,
\]
and the form factor becomes
\begin{align}
F_{\nu=3/5}(\widetilde{\mathbf{G}}) &=
\Bigl[1
+ \exp\!\Bigl(-\,i\,\tfrac{2\pi}{5}(i_1 + i_2)\,\Bigr)
+ \exp\!\Bigl(-\,i\,\tfrac{8\pi}{5}(i_1 + i_2),\Bigr)\Bigr]\,.
\end{align}

At $\nu=2/5$, there are two electrons in the unit cell at
\[
\{\mathbf{R}_M^{1,2}\}
=\Bigl\{\bigr(2a,\,0\bigr),
\bigl(3a,\,0\bigr)\Bigr\}\,,
\]
and the form factor becomes
\begin{align}
F_{\nu=2/5}(\widetilde{\mathbf{G}}) &=
\Bigl[\exp\!\Bigl(-\,i\,\tfrac{4\pi}{5}(i_1 + i_2)\Bigr)
+ \exp\!\Bigl(-\,i\,\tfrac{6\pi}{5}(i_1 + i_2)\Bigr)\Bigr].
\end{align}
The reciprocal lattice vectors in both $\nu=2/5$ and $\nu=3/5$ are 
\begin{equation}
\mathbf{\widetilde{G}} \;=\; i_1\,\mathbf{{G}}_1 + i_2\,\mathbf{{G}}_2 
\;=\;
\frac{2\pi}{a}\Bigl(\tfrac{i_1+i_2}{5} \;,\; \tfrac{-i_1 + i_2}{\sqrt3}\Bigr)\\[15pt].
\end{equation}

\subsubsection*{\textbf{Filling factor {$\nu = 6/7$}}}

At $\nu=6/7$, there are six electrons in the unit cell at
\[
\{\mathbf{R}_M^{1,2,3,4,5,6}\}
\;=\;
\Bigl\{
\Bigl(\tfrac{a}{2},\,\tfrac{\sqrt{3}\,a}{2}\Bigr),
\;(a,\,\sqrt{3}\,a),
\Bigl(\tfrac{3a}{2},\,\tfrac{3\sqrt{3}\,a}{2}\Bigr),
\Bigl(\tfrac{3a}{2},\,\tfrac{\sqrt{3}\,a}{2}\Bigr),
\;\bigl(2a,\,\sqrt{3}\,a\bigr),
\;\Bigl(\tfrac{5a}{2},\,\tfrac{3\sqrt{3}\,a}{2}\Bigr)
\Bigr\}.
\]
The form factor becomes
\begin{align}
F_{\nu=6/7}(\widetilde{\mathbf{G}}) &=
\Bigl[\exp\!\Bigl(-\,i\,\tfrac{2\pi}{7}(i_1 + 2i_2)\Bigr)
+ \exp\!\Bigl(-\,i\,\tfrac{2\pi}{7}(2i_1 + 4i_2)\Bigr)
+ \exp\!\Bigl(-\,i\,\tfrac{2\pi}{7}(3i_1 + 6i_2)\Bigr)\nonumber
\\[6pt]
&\quad
+ \exp\!\Bigl(-\,i\,\tfrac{2\pi}{7}(4i_1 + 1i_2)\Bigr)
+ \exp\!\Bigl(-\,i\,\tfrac{2\pi}{7}(5i_1 + 3i_2)\Bigr)
+ \exp\!\Bigl(-\,i\,\tfrac{2\pi}{7}(6i_1 + 5i_2)\Bigr)\Bigr],
\end{align}
where
\begin{equation}
\mathbf{\widetilde{G}} \;=\; i_1\,\mathbf{{G}}_1 + i_2\,\mathbf{{G}}_2 
\;=\;
\frac{2\pi}{a}\Bigl(\tfrac{3i_1-i_2}{7} \;,\; \tfrac{-i_1 + 5i_2}{7\sqrt3}\Bigr)\\[15pt].
\end{equation}

\subsubsection*{\textbf{Filling factor {$\nu \sim 1$}}}
At $\nu = 1 - \tfrac{1}{J^2}$ close to 1 ($J \geq 3$), there are ($J^2-1$)  electrons in the unit cell at 
\[
\{\mathbf{R}_M^{n,m}\}
\;=\;
\Bigl\{
\alpha_n\,\mathbf{R}_1
  \;+\;
  \beta_m\,\mathbf{R}_2\Bigl\},
\]

where $\alpha_n =\tfrac{n}{J}$ and $\beta_m = \tfrac{m}{J}$ (for integers $n, m$ with $0 \le n,m < J$), so that $0 \le \alpha_n, \beta_m < 1$. 
The form factor becomes 
\begin{align}
F_{\nu \sim 1}(\widetilde{\mathbf{G}}) 
&=
  -1 + \sum_{\substack{n,m}}^{}
  \exp\!\Bigl[
    -\,i\,\widetilde{\mathbf{G}}\cdot 
    \bigl(\alpha_n\,\mathbf{R}_1 + \beta_m\,\mathbf{R}_2\bigr)
  \Bigr].
\end{align}
The $-1$ term arises from the fact that there are ($J^2-1$) electrons in the moir\'{e} superlattice, and the term corresponding to $(n,m)=(0,0)$ has to be excluded.

\section{Matrix Elements $V_{\alpha,\beta}(\widetilde{\mathbf{G}})$}

In this section, we calculate the matrix elements, $V_{\alpha,\beta}(\widetilde{\mathbf{G}})$\;=\;$\langle \alpha \,\vert\, V(\widetilde{\mathbf{G}},\mathbf{r}) \,\vert\, \beta\rangle$, where \(\alpha,\beta \in \{2s, 2p_{+}, 2p_{-}\}\). These matrix elements are central to understanding the moir\'{e}-induced coupling between exciton states. From the main paper, the matrix elements is represented by
\begin{eqnarray}
V_{\alpha,\beta}(\widetilde{\mathbf{G}}) &=& \frac{2\pi e^2}{A_0\epsilon} \frac{e^{-d\widetilde{G}}}{\widetilde{G}}
\Bigg[ 2i F_{\nu}(\widetilde{\mathbf{G}})\Bigg]
\mathcal{M}_{\alpha,\beta}(\widetilde{\mathbf{G}}),
\end{eqnarray}
where $\widetilde{\mathbf{G}} \equiv \mathbf{G}-\mathbf{G}'$, and the function 
\begin{eqnarray}
\mathcal{M}_{\alpha,\beta}(\widetilde{\mathbf{G}}) = \int  \phi_\alpha^\ast(\mathbf{r}) \,\sin\Big(\tfrac{1}{2} \widetilde{\mathbf{G}}\cdot\mathbf{r} \Big) \,\phi_\beta(\mathbf{r}) \, d^2 \mathbf{r},
\end{eqnarray}
captures the coupling of the underlying hydrogen-like exciton wavefunctions $\phi_\alpha$ and $\phi_\beta$ in the presence of the sinusoidal factor. The matrix elements can also be written as
\begin{eqnarray}
V_{\alpha,\beta}(\widetilde{\mathbf{G}}) &=& \frac{2\pi e^2}{A_0\epsilon} \frac{e^{-d\widetilde{G}}}{\widetilde{G}}
\Bigg[F_{\nu}(\widetilde{\mathbf{G}})\Bigg]
\int  \phi_\alpha^\ast(\mathbf{r}) \,\Big(e^{\tfrac{1}{2}\,i \widetilde{\mathbf{G}}\cdot\mathbf{r}}-e^{-\tfrac{1}{2}\,i \widetilde{\mathbf{G}}\cdot\mathbf{r}}\Big) \,\phi_\beta(\mathbf{r}) \, d^2 \mathbf{r}.
\end{eqnarray}

\subsection{Representation of hydrogen-like states}

We consider the 2D hydrogen-like exciton states in an effective dielectric medium. In polar coordinates $(r,\theta)$, each eigenfunction can be written in the separable form:

\begin{equation}
\label{eq:phi}
\phi_{n,l}(r,\theta)
\;=\;
\frac{1}{\sqrt{2\pi}}\, e^{\,i\,l\,\theta}\, R_{n,l}(r),
\end{equation}
where $n$ is the principal quantum number and $l$ is the angular quantum number, taking values $l = 0,\pm 1,\pm 2, \dots, \pm(n - 1)$. The radial functions of 2$s$ and 2$p$ states are 
\begin{align}
R_{2s}(r) 
\; &=\; 
\frac{1}{\sqrt{3}}\,
\frac{(r_0 - r)}{r_0^2}
\, e^{-\tfrac{r}{2 r_0}}\nonumber,
\\[6pt]
R_{2p}(r)
&=\;
\frac{r}{\sqrt{6}\,r_0^2}
\, e^{-\tfrac{r}{2 r_0}}.
\end{align}
where $r_0 = \frac{3\,\epsilon\,\hbar^2}{4\,\mu\,e^2}$, $\mu$ is the reduced mass, $e$ is the electron charge, $\hbar$ is the reduced Planck constant, and $\epsilon$ is the dielectric constant ~\cite{S1}. Finally, we write the wavefunctions of the 2s and 2p states as:
\begin{align}
\phi_{2s}(r)
&=\;
\frac{1}{\sqrt{6\pi}}\,
\frac{(r_0 - r)}{r_0^2}
\, e^{-\tfrac{r}{2 r_0}}\nonumber,
\\[6pt] 
\phi_{2p\pm}(r,\theta)
&=\;
\frac{e^{\pm i\,\theta}}{\sqrt{12\pi}}\,
\frac{r}{r_0^2}\,
e^{-\tfrac{r}{2 r_0}}.
\end{align}

\subsection{Diagonal elements: \(\boldsymbol{V_{2s,2s}}\) and \(\boldsymbol{V_{2p_{\pm},2p_{\pm}}}\)}
The diagonal terms vanish
\begin{equation}
V_{2s,2s}(\widetilde{\mathbf{G}}) \;=\; 
\frac{2\pi e^2}{A_0\epsilon} \frac{e^{-d\widetilde{G}}}{\widetilde{G}}
\Bigg[ 2i F_{\nu}(\widetilde{\mathbf{G}})\Bigg]
\biggl(\int_{0}^{\infty} \Bigl[\int_{0}^{2\pi} \sin\Big(\tfrac{1}{2} \widetilde{\mathbf{G}}\cdot\mathbf{r}\Bigr)\,\frac{1}{2\pi}\,d\theta\Bigr]\
R_{2s}^{2}(r)\,dr\biggr)\;=\;0,
\end{equation}

\begin{equation}
V_{2p_{\pm},2p_{\pm}}(\widetilde{\mathbf{G}}) \;=\;\frac{2\pi e^2}{A_0\epsilon} \frac{e^{-d\widetilde{G}}}{\widetilde{G}}
\Bigg[ 2i F_{\nu}(\widetilde{\mathbf{G}})\Bigg]
\biggl(\int_{0}^{\infty} \Bigl[\int_{0}^{2\pi} \sin\Big(\tfrac{1}{2} \widetilde{\mathbf{G}}\cdot\mathbf{r}\Bigr)\,\frac{1}{2\pi}\,d\theta\Bigr]\,
R_{2p}^{2}(r)\,dr\biggr)\;=\;0,
\end{equation}
\\
where $\mathbf{r}=(r,\theta)$ and $\tilde{\mathbf{G}}=(\tilde{G},\varphi)$ in polar coordinates.
\\[6pt]

\subsection{Off-diagonal elements: \(\boldsymbol{V_{2s,2p_{\pm}}}\), \(\boldsymbol{V_{2p_{\pm},2s}}\), and \(\boldsymbol{V_{2p_{\mp},2p_{\pm}}}\)}

Next, we calculate the off-diagonal terms. To present the calculation steps more clearly, we replace the sinusoidal term with its exponential form in Eq.~(S15), and get
\begin{equation}
\begin{aligned}
V_{2s,2p_{+}}(\widetilde{\mathbf{G}})
&=\frac{2\pi e^2}{A_0\epsilon} \frac{e^{-d\widetilde{G}}}{\widetilde{G}}
\Bigg[ F_{\nu}({\widetilde{\mathbf{G}}})\Bigg]\int_{0}^{\infty}\int_{0}^{2\pi}
\Bigl(e^{\,i\,{\widetilde{G}\,r\,\cos(\theta-\varphi)/2}} \;-\; e^{-\,i\,{\widetilde{G}}\,r\,\cos(\theta-\varphi)/2}\Bigr)\,
\frac{1}{2\pi}\,e^{\,i\,\theta}\,d\theta\;R_{2s}(r)\,R_{2p}(r)\,r\,dr\\
&=\ [\dots] \int_{0}^{\infty} e^{\,i\,\varphi}\,
\int_{0}^{2\pi}
\Bigl(e^{\,i\,\tfrac{\widetilde{G}\,r}{2}\cos\theta} - e^{-\,i\,\tfrac{\widetilde{G}\,r}{2}\cos\theta}\Bigr)\,
\frac{1}{2\pi}\,e^{\,i\,\theta}\,d\theta\;R_{2s}(r)\,R_{2p}(r)\,r\,dr\\
&=\ [\dots] e^{\,i\,\varphi}\,i
\int_{0}^{\infty}
\Bigl[
  J_{1}\!\Bigl(\tfrac{\widetilde{G}\,r}{2}\Bigr)
  \;-\;
  J_{-1}\!\Bigl(\tfrac{-\widetilde{G}\,r}{2}\Bigr)
\Bigr]
\,R_{2s}(r)\,R_{2p}(r)\,r
\,dr\\
&=\ [\dots] 
2\,i\,e^{\,i\,\varphi}\
\int_{0}^{\infty}
J_{1}\!\Bigl(\tfrac{\widetilde{G}\,r}{2}\Bigr)\,
R_{2s}(r)\,R_{2p}(r)\,r
\,dr\\
&=\; [\dots] 2\,i\,e^{\,i\,\varphi}\,\frac{1}{3\sqrt{2}r_0^3}\ \int_{0}^{\infty} r^2\,(1 - r/r_0)\,J_{1}\!\Bigl(\tfrac{\widetilde{G}\,r}{2}\Bigr)\,e^{-r/r_0}\,dr\\
&=\frac{2\pi e^2}{A_0\epsilon} \frac{e^{-d\widetilde{G}}}{\widetilde{G}}
\Bigg[ 2i F_{\nu}(\widetilde{\mathbf{G}})\Bigg] \times
16\,\sqrt{2}\,e^{\,i\,\varphi}\;
\frac{
  (r_0^2\widetilde{G}^{2} - 6\bigr)r_0\widetilde{G}
}{
  \bigl(r_0^2\widetilde{G}^{2} + 4\bigr)^{7/2}
}.
\end{aligned}
\end{equation}
where $r_{0}=3\epsilon\hbar^2 / 4\mu e^2 $. Similarly, one can show  that
\begin{equation}
\begin{aligned}
V_{2s,2p_{-}}(\widetilde{\mathbf{G}})
&=\frac{2\pi e^2}{A_0\epsilon} \frac{e^{-d\widetilde{G}}}{\widetilde{G}}
\Bigg[ F_{\nu}(\widetilde{\mathbf{{G}}})\Bigg]\int_{0}^{\infty}\int_{0}^{2\pi}
\Bigl(e^{\,i\,\widetilde{G}\,r\,\cos(\theta-\varphi)/2} \;-\; e^{-\,i\,\widetilde{G}\,r\,\cos(\theta-\varphi)/2}\Bigr)\,
\frac{1}{2\pi}\,e^{\,-i\,\theta}\,d\theta\;R_{2s}(r)\,R_{2p}(r)\,r\,dr\\
&=\ [\dots] \int_{0}^{\infty} e^{-\,i\,\varphi}\,
\int_{0}^{2\pi}
\Bigl(e^{\,i\,\tfrac{\widetilde{G}\,r}{2}\cos\theta} - e^{-\,i\,\tfrac{\widetilde{G}\,r}{2}\cos\theta}\Bigr)\,
\frac{1}{2\pi}\,e^{-\,i\,\theta}\,d\theta\;R_{2s}(r)\,R_{2p}(r)\,r\,dr\\
&=\ [\dots] e^{-\,i\,\varphi}\,i^{-1}
\int_{0}^{\infty}
\Bigl[
  J_{-1}\!\Bigl(\tfrac{\widetilde{G}\,r}{2}\Bigr)
  \;-\;
  J_{-1}\!\Bigl(\tfrac{-\widetilde{G}\,r}{2}\Bigr)
\Bigr]
\,R_{2s}(r)\,R_{2p}(r)\,r
\,dr\\
&=\ [\dots] 
2\,i\,e^{-\,i\,\varphi}\
\int_{0}^{\infty}
J_{1}\!\Bigl(\tfrac{\widetilde{G}\,r}{2}\Bigr)\,
R_{2s}(r)\,R_{2p}(r)\,r
\,dr\\
&=\; [\dots] 2\,i\,e^{-\,i\,\varphi}\,\frac{1}{3\sqrt{2}r_0^3}\ \int_{0}^{\infty} r^2\,(1 - r/r_0)\,J_{1}\!\Bigl(\tfrac{\widetilde{G}\,r}{2}\Bigr)\,e^{-r/r_0}\,dr\\
&=\frac{2\pi e^2}{A_0\epsilon} \frac{e^{-d\widetilde{G}}}{\widetilde{G}}
\Bigg[2iF_{\nu}(\widetilde{\mathbf{G}})\Bigg] \times
16\,\sqrt{2}\,e^{-\,i\,\varphi}\;
\frac{
  (r_0^2\widetilde{G}^{2} - 6\bigr)r_0\widetilde{G}
}{
  \bigl(r_0^2\widetilde{G}^{2} + 4\bigr)^{7/2}
}.
\end{aligned}
\end{equation}

\vspace{2mm}

The matrix elements in Eqs.~(S21) and (S22) differ only by a phase factor $e^{\,i\,\varphi}$, consistent with the requirement that the overall Hamiltonian is Hermitian. Namely,
\[
V_{2s,2p_{-}}(\widetilde{\mathbf{G}})
\;=\;
V_{2p_{+},2s}(\,\widetilde{\mathbf{G}})
\;=\;
\bigl[V_{2p_{-},2s}(\,-\widetilde{\mathbf{G}})\bigr]^*
\;=\;
\bigl[V_{2s,2p_{+}}(\,-\widetilde{\mathbf{G}})\bigr]^*\,\,.
\]

\vspace{2mm}

The zero coupling also arises between the states \(2p_{\pm}\) and \(2p_{\mp}\)
\begin{equation}
\begin{aligned}
V_{2p_{\pm},2p_{\mp}}(\widetilde{\mathbf{G}})
&=\frac{2\pi e^2}{A_0\epsilon} \frac{e^{-d\widetilde{G}}}{\widetilde{G}}
\Bigg[ F_{\nu}(\widetilde{\mathbf{{G}}})\Bigg]\int_{0}^{\infty}\int_{0}^{2\pi}
\Bigl(e^{\,i\,\widetilde{G}\,r\,\cos(\theta-\varphi)/2} \;-\; e^{-\,i\,\widetilde{G}\,r\,\cos(\theta-\varphi)/2}\Bigr)\,
\frac{1}{2\pi}\,e^{\,2i\,\theta}\,d\theta\;R_{2p}(r)\,R_{2p}(r)\,r\,dr\\
&=\ [\dots] \int_{0}^{\infty} e^{2\,i\,\varphi}\,
\int_{0}^{2\pi}
\Bigl(e^{\,i\,\tfrac{\widetilde{G}\,r}{2}\cos\theta} - e^{-\,i\,\tfrac{\widetilde{G}\,r}{2}\cos\theta}\Bigr)\,
\frac{1}{2\pi}\,e^{2\,i\,\theta}\,d\theta\;R_{2p}(r)\,R_{2p}(r)\,r\,dr\\
&=\ [\dots] e^{2\,i\,\varphi}\,i^{2}
\int_{0}^{\infty}
\Bigl[
  J_{2}\!\Bigl(\tfrac{\widetilde{G}\,r}{2}\Bigr)
  \;-\;
  J_{2}\!\Bigl(\tfrac{-\widetilde{G}\,r}{2}\Bigr)
\Bigr]
\,R_{2p}(r)\,R_{2p}(r)\,r
\,dr \,\, =\,\,0\,\,.
\end{aligned}
\end{equation}

\subsection{Summary of final results}
In conclusion, we have obtained the following matrix elements. The diagonal elements, as well as the mixtures between $2p_{+}$ and $2p_{-}$, vanish
\begin{eqnarray}
V_{2s,2s}(\widetilde{\mathbf{G}}) = 0,\quad
V_{2p_{\pm},2p_{\pm}}(\widetilde{\mathbf{G}}) = 0,\quad
V_{2p_{\pm},2p_{\mp}}(\widetilde{\mathbf{G}}) = 0\,\,,
\end{eqnarray}
while the off-diagonal terms $V_{2s,2p_{\pm}}(\mathbf{G})$ and their Hermitian conjugates are nonzero and yield
\begin{eqnarray}
V_{2s,2p_{+}}(\widetilde{\mathbf{G}}) \;=\; M_{+}(\widetilde{\mathbf{G}}), 
\quad
V_{2s,2p_{-}}(\widetilde{\mathbf{G}}) \;=\; M_{-}(\widetilde{\mathbf{G}}),
\quad
V_{2p_{+},2s}(\widetilde{\mathbf{G}}) \;=\; M_{-}(\widetilde{\mathbf{G}}), 
\quad
V_{2p_{-},2s}(\widetilde{\mathbf{G}}) \;=\; M_{+}(\widetilde{\mathbf{G}})\,\,,
\end{eqnarray}
where
\begin{equation}
M_{\pm}(\widetilde{\mathbf{G}})
=\frac{2\pi e^2}{A_0\epsilon} \frac{e^{-d\widetilde{G}}}{\widetilde{G}}
\Bigg[ 2i F_{\nu}(\widetilde{\mathbf{G}})\Bigg] \times
16\,\sqrt{2}\,e^{\,\pm i\,\varphi}\;
\frac{
  (r_0^2\widetilde{G}^{2} - 6\bigr)r_0\widetilde{G}
}{
  \bigl(r_0^2\widetilde{G}^{2} + 4\bigr)^{7/2}
}\,\,\,. \nonumber
\end{equation}
These expressions confirm that only off-diagonal terms that couple 2$s$ and 2$p$ orbitals survive, and capture how the factors $\widetilde{{G}}$ and \(e^{\pm i\,\varphi}\) of $\widetilde{\mathbf{G}}$ appear naturally from integration. 

\section{Disordered Configuration}


The main idea is illustrated schematically in Fig. S3(a), where each filled circle represents a site that is occupied by a charge. To investigate the effects of disorder, we first define a supercell made of  $N\times N$ basic cells and randomly distribute the electrons in the supercell according to the filling factor. Figure S3(a) show an example when $N=12$ and $\nu\,$$=$$\,$0.5, in which case the total number of electrons in the supercell is $\nu N^2=72$.

\vspace{3mm}

Making use of the lattice vectors of $\nu\,$$=$$\,$1, the lattice vectors of the large supercell can be defined by
\begin{equation}
\begin{aligned}
\mathbf{R}_1 &= a \,(1, 0), \\
\mathbf{R}_2 &= \tfrac{a}{2}\,(1, \sqrt{3}), \\
\mathbf{G}_1 &= \tfrac{2\pi}{a}\,\bigl(1, -\tfrac{1}{\sqrt{3}}\bigr), \\
\mathbf{G}_2 &= \tfrac{4\pi}{a}\,\bigl(0, \tfrac{1}{\sqrt{3}}\bigr),
\end{aligned}
\quad
\Longrightarrow
\quad
\begin{aligned}
\mathbf{R}_1^{\mathrm{large}} &= aN \,(1, 0), \\
\mathbf{R}_2^{\mathrm{large}} &= \tfrac{aN}{2}\,(1, \sqrt{3}), \\
\mathbf{G}_1^{\mathrm{large}} &= \tfrac{2\pi}{aN}\,\bigl(1, -\tfrac{1}{\sqrt{3}}\bigr), \\
\mathbf{G}_2^{\mathrm{large}} &= \tfrac{4\pi}{aN}\,\bigl(0, \tfrac{1}{\sqrt{3}}\bigr).
\end{aligned}
\end{equation}

A general site in the supercell, whether occupied or empty, can be written as
\[
  \mathbf{R}_{n,m}
  \;=\;
  \alpha_n\,\mathbf{R}_1
  \;+\;
  \beta_m\,\mathbf{R}_2,
\]
where $\alpha_n = n/N$ and $\beta_m = m/N$ are fractions in the interval $0 \leq \alpha_n,\beta_m < 1$, where $n$ and $m$ are integers in the range $0 \le i,j < N$. The form factor becomes
\[
   F_{\nu}(\widetilde{\mathbf{G}}) \;\equiv\; \sum_{ \mathbf{R}_{n,m}'} e^{-\,i\,\widetilde{\mathbf{G}}\cdot  \mathbf{R}_{n,m}'} \,\,\,\,,
\]
where the prime in $\mathbf{R}'_{n,m}$ signifies that the sum only runs over occupied sites.
%
Because the distribution of occupied sites can be fully random, the phase factors $e^{-i\,\tilde{\mathbf{G}}\cdot\mathbf{R}_{n,m}'}$ will not in general sum constructively. Consequently, the magnitude of $F_{\nu}(\widetilde{\mathbf{G}})$ can vary significantly from sample to sample.

\subsection{Result of optical absorption by disordered states}
\begin{figure}[h]
    \centering
    \includegraphics[width=0.95\textwidth]{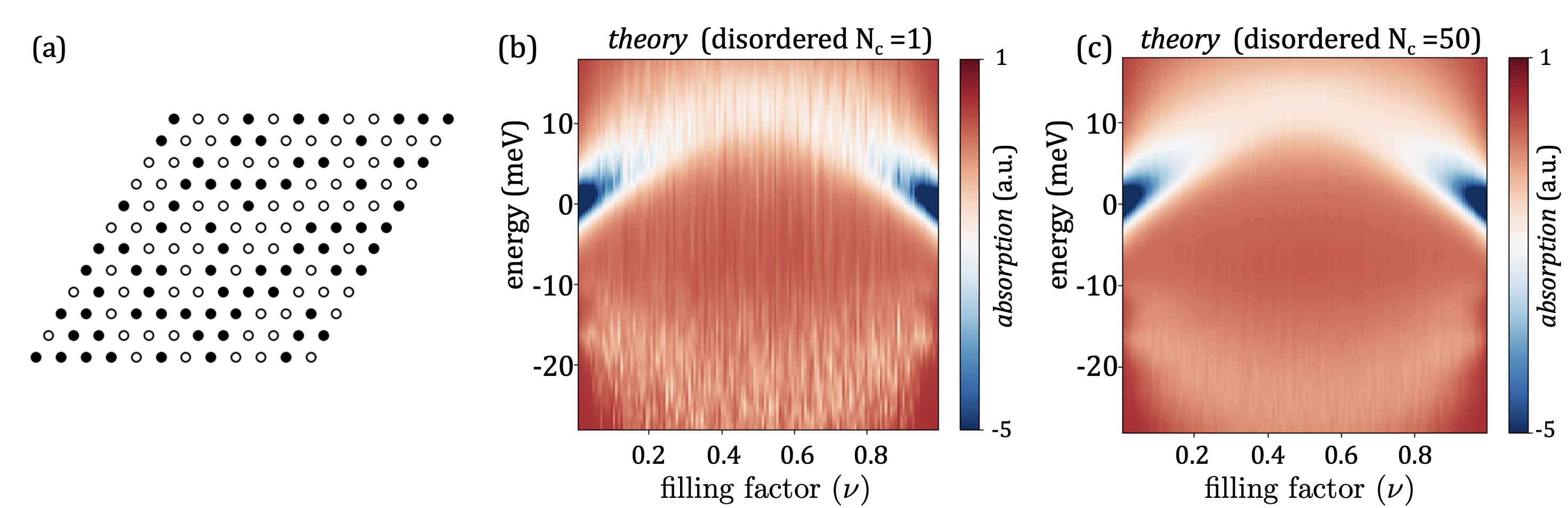}
    \caption{\label{fig:S3}
        (a) Example of a random distribution used to compute the absorption in (b) for a $12\times12$ extended supercell, where $72$ out of $144$ sites are occupied, corresponding to $\nu = 0.5$. 
        (b) Optical absorption spectrum for one random configuration ($N_c = 1$).
        (c) Same as (b) but averaged over $50$ random configurations ($N_c = 50$).
    }
\end{figure}
Figure~S3(a) shows an example of a random distribution whose absorption profile is shown in Fig. S3(b). In this illustration, there are $12\times12 = 144$ lattice sites in the supercell, and half of them ($72$ sites) are occupied by electrons, corresponding to a filling factor $\nu = 0.5$. Unlike a moir\'{e} superlattice with well-defined periodic potentials, these charges are placed at random positions.

\vspace{2mm}

To quantify how such randomness affects the optical spectrum, we have computed the absorption under two scenarios for each filling factors \(\nu\).  Figures~S3(b) and (c) correspond to a \(12\times12\) supercell but differ in the number of disorder realizations considered for each filling factor. Specifically, Fig.~S3(b) shows the absorption when only a single random sample is used for each \(\nu\), while Fig.~S3(c) shows the absorption averaged over 50 different random configurations. Despite minor speckling arising from the randomness in the single-sample data [Fig.~S3(b)], the overall profile is already similar to the averaged result [Fig.~S3(c)].

\vspace{2mm}

These results confirm that the distinct resonance features observed experimentally fail to emerge in the simulation if we employ a disordered configuration, and furthermore, it is not merely an artifact of averaging over multiple disorder realizations. These simulations reaffirm that charge-ordered states in the moir\'{e} superlattice are responsible for the results seen in experiment, as we show in Fig~2(b) of the main paper.

\subsection{Numerical aspects}
The size of the matrix diagonalize in Eq. (8) of the main paper is  $\mathcal{N}\times \mathcal {N}$ where $\mathcal{N}=3(2N_G+1)^2$. The factor 3 comes from the number of exciton states we employ ($2s$ and $2p^{\pm}$), and the factor $(2N_G+1)^2$ corresponds to the number of reciprocal lattice vectors used in the simulations. In more detail, the reciprocal lattice vectors are $\widetilde{\mathbf{G}} = i_1\,\mathbf{G}_1 + i_2\,\mathbf{G}_2$, where $i_1$ and $i_2$ are integers in the interval $[-N_G,\,N_G]$. Hence, there are $(2N_G+1)^2$ possible combinations of $i_1$ and $i_2$. We have found that it is sufficient to use $N_G=8$ when simulating ordered configurations ($\mathcal {N} = 867$), and  $N_G=13$ when simulating disordered configurations ($\mathcal {N} = 2187$). In the disordered case, the reciprocal lattice vectors are $\widetilde{\mathbf{G}} = i_1\,\mathbf{G}_1^{\mathrm{large}} + i_2\,\mathbf{G}_2^{\mathrm{large}}$, where $\mathbf{G}_{1(2)}^{\mathrm{large}}$ were defined in Eq.~(S28). Given that all of the matrix elements are analytical, the main numerical task is to diagonalize the matrix and not to compute its elements. Since $\mathcal{N}$ is of the order of 1000, the computation time for a given filling factor is of the order of minutes on a simple laptop.

\end{widetext}

\end{document}